\documentclass[11pt]{article}
\usepackage{amsmath,amsfonts}
\usepackage{amssymb}
\usepackage{fullpage}
\usepackage{lscape}
\usepackage[dvips]{epsfig}

\def\##1{\underline #1}
\def\=#1{\underline{\underline #1}}

\def\eps{\epsilon}

\def\lambdao{\lambda_0}
\def\etao{\eta_0}
\def\.{\mbox{ \tiny{$^\bullet$} }}

\def\epsaj{\epsilon_{a_j}}
\def\epsbj{\epsilon_{b_j}}
\def\epscj{\epsilon_{c_j}}
\def\epsdj{\tilde{\epsilon}_{d_j}}

\def\ux{{ \#u}_x}
\def\uy{{ \#u}_y}
\def\uz{{ \#u}_z}

\def\Sz{{ \=S}(z)}

\def\le{\left(}
\def\ri{\right)}
\def\les{\left[}
\def\ris{\right]}
\def\lec{\left\{}

\def\c#1{\cite{#1}}

\def\r#1{(\ref{#1})}

\begin{document}

\begin{center}

{\bf \large Third method for generation of spectral holes \\ in
chiral sculptured thin films}\\
\bigskip

Fei Wang\footnote{Corresponding Author. Fax: +1 814 863 7967;
e--mail: fuw101@psu.edu} and Akhlesh Lakhtakia\footnote{E--mail:
akhlesh@psu.edu}\\

{\em CATMAS --- Computational and Theoretical Materials Sciences Group \\
     Department of Engineering Science and Mechanics \\
     Pennsylvania State University, University Park, PA
     16802--6812, USA}

\end{center}


\noindent {\bf Abstract:} The introduction of either a central layer defect or a central
twist defect in a periodic structurally chiral material generates circular--polarization--sensitive spectral holes
in the remittance spectrums for normally incident plane waves. We propose and theoretically establish here the third method  to generate such
spectral holes using two--section chiral sculptured thin films (STFs).  Both sections of the proposed device
have the same periodicity and handedness, but their dielectric properties are different and related in a specific way.
The concept of pseudoisotropy is highly relevant for the production of the proposed device.


\section{Introduction}
The generation of an intra--band spectral hole was first
demonstrated by inserting a phase defect in the center of a scalar Bragg grating  \c{Haus}. The
scalar Bragg grating, without the central phase defect, has a spectral regime of high
reflectance for normally incident plane waves. This regime is called the Bragg regime. When the central phase
defect is inserted, the Bragg regime is punctured by a
much narrower high--transmittance regime. This second regime is
called a spectral (reflection) hole and is widely employed in
laser optics \c{Agra1} as well as in optical--fiber communication
\c{Agra2, Bak}.

As a scalar Bragg grating is insensitive to the polarization state
of the incident plane wave, the incorporation of a central phase defect gives rise to a reflection hole
regardless of the polarization state. In order to generate
circular--polarization--sensitive reflection holes,  periodic structurally
chiral materials~---~exemplified by chiral sculptured thin films (STFs)
and cholesteric liquid crystals (CLCs) \c{AL99}--\c{Kopp3}~---~are
used in lieu of scalar Bragg gratings. In general, these materials
discriminate between incident plane waves of different circular
polarization states in the Bragg regime. Periodic structurally chiral materials and circularly polarized plane waves
possess handedness.
In the Bragg regime,
the reflectance is very high
for a co--handed normally incident plane wave, but not for the
cross--handed one~---~leading to the term {\em circular Bragg phenomenon\/}.
As the high reflectance in the Bragg regime is
co--handed only, so is the reflection hole in the Bragg regime
generated by the insertion of a central phase defect in the periodic structurally chiral
material.

Theoretical analysis has recently engendered another
spectral hole~---~i.e., a cross--handed transmission
hole~---~in a periodic structurally chiral material
by the introduction of a central phase defect \c{Kopp1}--\c{Hodg}.
The thickness of the periodic structurally chiral material is
crucial to the exhibition  of the two types of spectral
holes. When the thickness is relatively small, the co--handed
reflection hole occurs. As the thickness increases, the co--handed
reflection hole diminishes to vanish eventually and it is gradually
replaced by the cross--handed transmission hole. The bandwidth of the second
type of spectral holes is a tiny fraction of that of the first type. However, even modest dissipation can be deleterious
to the second type of spectral holes \c{FW1}.
Needless to add, the second type cannot be generated
using scalar Bragg gratings.

The central phase defects investigated thus far are of two types:
\begin{itemize}
\item[(i)] Layer defect: A homogeneous layer, whether isotropic \c{AL99,HH1} or anisotropic \c{LVM},
is inserted in the center of the grating. The thickness of the homogeneous layer determines the
center wavelength of the spectral hole, with a quarter--wave layer positioning the spectral hole quite accurately
in the center of the Bragg regime \c{LM05}.
\item[(ii)] Twist defect: One half of the periodic structurally chiral material is rotated about the thickness
axis with respect to the other half by a certain angle \c{Hodg3,Kopp1,FW1,Schmidtke1} The amount of rotation
determines the center--wavelength of the spectral hole, with a $90^\circ$--twist positioning
the spectral hole
in the center of the Bragg regime.
\end{itemize}
Combinations of the two types of phase defects are likely to offer superior performance
than either alone \c{Hodg}.

We propose here a third method to generate both types of spectral
holes in periodic structurally chiral materials. This method can
be implemented with chiral STFs but not with CLCs. It is based on
the selection of a two--section chiral STF with the two sections
having  different dielectric properties but the same periodicity
and the same handedness. The relevant  boundary value problem for
normally incident plane waves is briefly described in Section 2,
while the proposed method and its possible implementation are
examined in Section 3. The concept of pseudoisotropy is highy relevant 
to the implementation of our proposal \c{Abd,LW}.

\begin{figure}[!ht]
\centering \psfull \epsfig{file=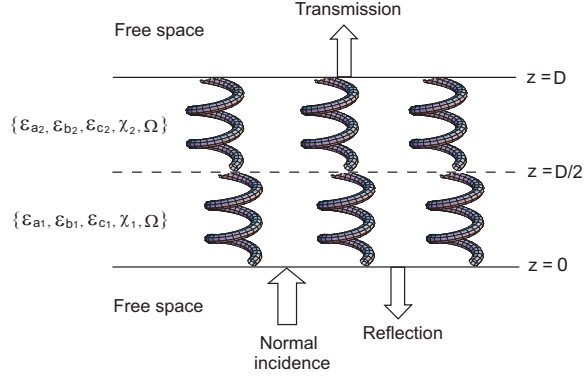,width=3in}
\caption{Schematic of the boundary value problem involving a
two--section chiral STF whose lower and upper sections have
different reference relative permittivity scalars $\eps_{a,b,c}$
and tilt angle $\chi$. The structural half--period $\Omega$ and
handedness are uniform throughout the chiral STF. }
\label{fig_Geom}
\end{figure}

\section{Theory}
Suppose the region $0<z<D$ is occupied by a two--section chiral
STF, while the half--spaces $z\leq 0$ and $z \geq D$ are vacuous,
as shown in Figure \ref{fig_Geom}. The relative permittivity
dyadic of the chiral STF is stated as follows:
\begin{equation}
\=\eps(\#r) =\lec \begin{array}{l} \Sz \cdot \=S_y(\chi_1) \cdot
\les \eps_{a_1} \uz\uz +\eps_{b_1} \ux\ux + \eps_{c_1} \uy\uy \ris
 \\ [5pt]{}\qquad\cdot
\=S_y^{-1}(\chi_1) \cdot \=S_z^{-1}(z)\,,\quad  0<z<D/2\,, \\[5pt]
\=S_z(z-D/2) \cdot \=S_y(\chi_2) \cdot \les \eps_{a_2} \uz\uz +\eps_{b_2}
\ux\ux + \eps_{c_2} \uy\uy \ris \\[5pt]{} \qquad\cdot \=S_y^{-1}(\chi_2) \cdot
\=S_z^{-1}(z-D/2)\,, \quad D/2<z<D\,. \end{array} \right.
\label{epsbasic}
\end{equation}
Here and hereafter, $\eps_{a_j, b_j, c_j}$, ($j=1,2$), are the
reference relative permittivity scalars of the $j$th section; $\{ \ux, \uy, \uz \}$ are the unit cartesian vectors with
$\uz$ parallel to the axis of nonhomogeneity of the chiral STF;
the rotational dyadic
\begin{equation}
\=S_z(z)= \uz\uz + \le \ux\ux+\uy\uy\ri \cos pz  + h \le
\uy\ux-\ux\uy\ri \sin pz\,,
\end{equation}
where  $p=\pi /\Omega$, $2\Omega$ is the structural period, and
the parameter $h=1$ for structural right--handedness and $h=-1$
for structural left--handedness; and the tilt dyadic
\begin{equation}
\=S_y(\chi_j) = \uy\uy + (\ux\ux + \uz\uz)  \cos\chi_j +
(\uz\ux-\ux\uz) \sin\chi_j\,, \quad  j=1, 2\,,
\end{equation}
represents the {\it locally\/} aciculate morphology of the STF
with $\chi_j$ as the tilt angle. The scalars $\eps_{a_j, b_j,
c_j}$, ($j=1,2$), are implicitly dependent on the free--space
wavelength $\lambda_0$,  and $\chi_j>0$ for chiral STFs. The ratio
$D/2\Omega$ is set as a positive integer. For convenience, we
define
\begin{equation}
\label{epsd} \epsdj = \frac{\epsaj \epsbj}{\epsaj \cos^2\chi_j +
\epsbj \sin^2\chi_j}\,, \quad  j=1, 2 \,.
\end{equation}
Parenthetically, the relative permittivity dyadic of CLCs
can be described by \r{epsbasic} as well, but with the
restrictions $\chi_j\equiv0$ and $\eps_{a_j}=\\equiv\eps_{c_j}$.

The two--section chiral STF is axially excited by a normally
incident, circularly polarized plane wave from the half--space
$z\leq 0$. The procedure to obtain the planewave reflectances and
transmittances is devised from the solution of a boundary value
problem detailed elsewhere \c{LM05,VL98a}. Let us content
ourselves here by stating that $4\times4$ algebraic matrix
equation \c{AL99}
\begin{equation}
\label{ame} [\#f_{exit}] = [\=M] [\#f_{entry}]
\end{equation}
eventually emerges, where the column--4 vectors $[\#f_{entry}]$
and $[\#f_{exit}]$ contain the $x$-- and the $y$-- components of
the electromagnetic field phasors at the entry and the exit
pupils, respectively. The $4\times$4 matrix
\begin{equation}
\label{meqn} [\=M] =  \exp \Big( i [\=P_2] D/2\Big)
  \exp \Big( i [\=P_1]D/2 \Big)  \,
\end{equation}
relating $[\#f_{entry}]$ and $[\#f_{exit}]$  is computed using the
matrixes
\begin{equation}
\label{pmat} [\=P_j] = \les \begin{array} {cccc}
0 & - ihp & 0 &\frac{2\pi\etao}{\lambdao}\\[3pt]
ihp & 0 & - \frac{2\pi\etao}{\lambdao} & 0\\[3pt]
0& -\frac{2\pi}{\lambdao\etao} \epscj& 0 &-  ihp\\[3pt]
\frac{2\pi}{\lambdao\etao} \epsdj& 0 & ihp & 0
\end{array}\ris \,, \quad j=1, 2\,,
\end{equation}
where $\etao$ is the intrinsic impedance of free space. The
derivation of \r{meqn} does not account for the possible
excitation of Voigt waves \c{a}; but that possibility is remote,
and can occur only for highly dissipative chiral STFs \c{b}.

\section{Proposed Method}

The planewave remittances (i.e., reflectances and transmittances) can be easily computed after solving
\r{ame}. But our interest was in finding relationships between
$\eps_{c_j}$ and $\epsdj$, ($j=1,2$) for the
generation of the two types of spectral holes.

We determined that the relationships
\begin{equation} \label{req} \lec \begin{array}{l}
\eps_{c_1}=\tilde{\eps}_{d_2} \\
\tilde{\eps}_{d_1}=\eps_{c_2} \end{array} \right.
\end{equation}
lead to the identity
\begin{equation}
[\=P_2]=[\=B(\pi/2)][\=P_1][\=B(\pi/2)]^{-1} \,.
\end{equation}
Therefore, satisfaction of the conditions \r{req} implies that \r{meqn} converts to
\begin{equation}
\label{meqn1} [\=M]=  [\=B(\pi/2)] \exp \Big(i [\=P_1]
D/2 \Big) [\=B(\pi/2)]^{-1}
  \exp \Big( i [\=P_1]D/2 \Big)\,,
  \end{equation}
  where
\begin{equation}
[\=B(\phi)] = \les \begin{array}{cccc}
\cos\phi & - h \sin\phi & 0 & 0\\
h \sin\phi & \cos\phi & 0 & 0\\
0 & 0 & \cos\phi & - h\sin\phi\\
0 & 0 & h \sin\phi & \cos\phi\end{array}\ris\,.
\end{equation}

\begin{figure}[!ht]
\centering \psfull \epsfig{file=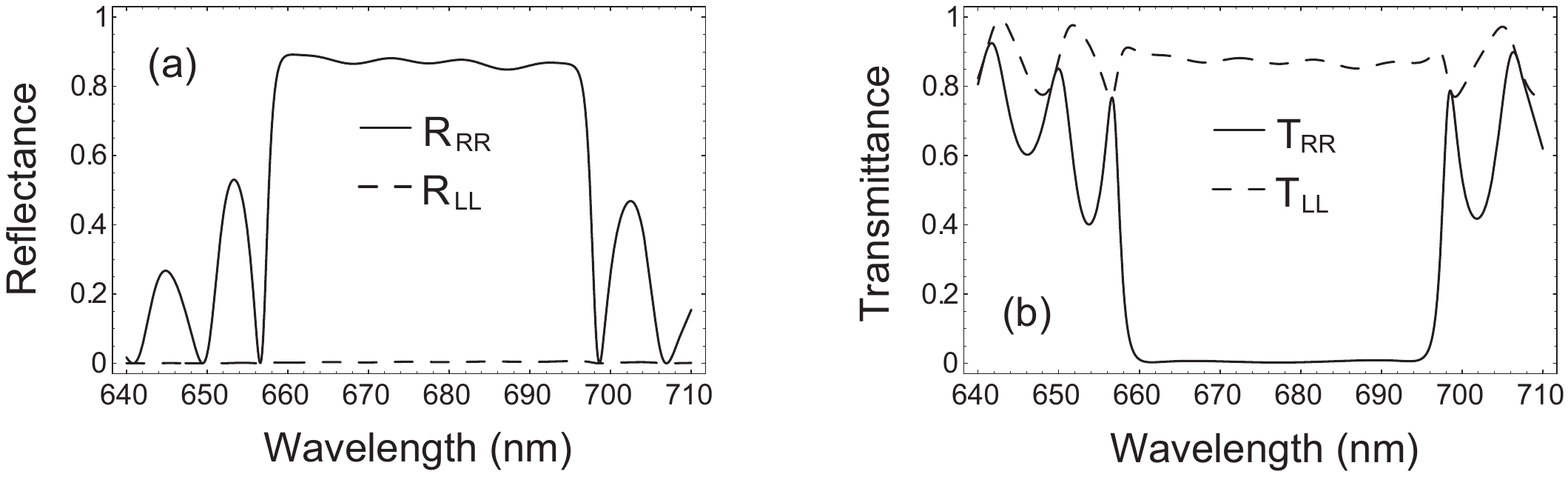,width=4in}
\caption{Spectrums of (a) reflectances ${\rm R}_{\rm RR}$ and
${\rm R}_{\rm LL}$ and (b) transmittances ${\rm T}_{\rm RR}$ and
${\rm T}_{\rm LL}$, computed for a structurally right--handed,
defect--free chiral STF with
 $\eps_{c_1}=\eps_{c_2}=2.72$,
$\tilde{\eps}_{d_1}=\tilde{\eps}_{d_2}=3.02$, $\Omega=200$ nm, and $D=60\, \Omega$. The
Bragg regime of the chiral STF is estimated as $600 < \lambda_0
<695$~nm. The circular Bragg phenomenon is evident as a high
co--handed reflectance (${\rm R}_{\rm RR}$) 
and a high cross--handed transmittance (${\rm T}_{\rm LL}$) in the Bragg regime. (The double subscript {\rm LL} in
${\rm T}_{\rm LL}$ indicates that the incident and the transmitted
plane waves are {\em L}eft circularly polarized. Likewise, ${\rm T}_{\rm
RR}$ is  the transmittance  of an incident {\em R}ight circularly
polarized plane wave as a {\em R}ight circularly polarized plane wave.
The cross--polarized remittances, such as ${\rm R}_{\rm LR}$,
etc., can be minimized by using index--matching layers \c{HH2},
which was not implemented for this figure.) } \label{Fig2}
\end{figure}

\begin{figure}[!ht]
\centering \psfull \epsfig{file=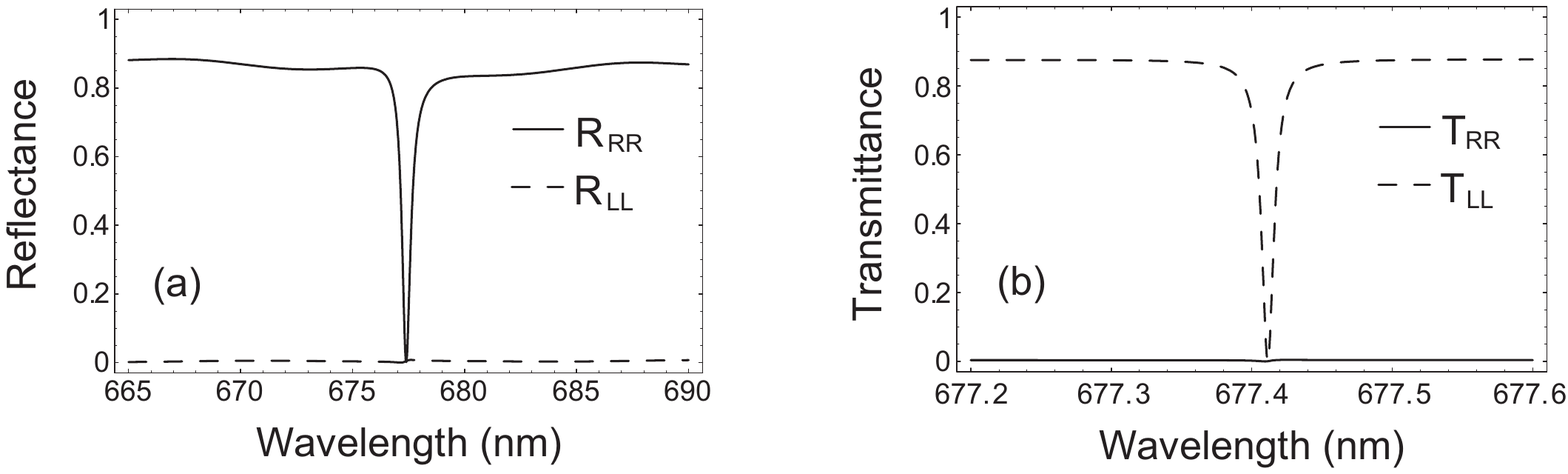,width=4in}
\caption{Spectrums of (a) reflectances ${\rm R}_{\rm RR}$ and
${\rm R}_{\rm LL}$ and (b) transmittances ${\rm T}_{\rm RR}$ and
${\rm T}_{\rm LL}$, computed for a structurally right--handed,
two--section chiral STF with
 $\eps_{c_1}=\tilde{\eps}_{d_2}=2.72$,
$\tilde{\eps}_{d_1}=\eps_{c_2}=3.02$, and $\Omega=200$ nm. The
thickness of the device is (a) $D=60 \, \Omega$ and (b) $D=180 \,
\Omega$. Compared with the remittance spectrums in Figure 2 for
the defect--free chiral STF, the remittance spectrums of the
two--section chiral STF exhibit two types of spectral holes in the
center of the Bragg regime.  A co--handed reflection hole in the
spectrum of ${\rm R}_{\rm RR}$ emerges in the center of the Bragg
regime when $D$ is relatively small. As $D$ increases, the
co--handed reflection hole vanishes and is replaced by a
cross--handed transmission hole in the spectrum of ${\rm T} _{\rm
LL}$ when $D$ is sufficiently large. } \label{Fig3}
\end{figure}

The matrix $[\=M]$ of \r{meqn1} is identical to that formulated
for an axially excited chiral STF with a central $90^\circ$--twist
defect \c{Hodg3}. Therefore, the two--section chiral STF
satisfying the conditions \r{req} should resemble a chiral STF
with a central $90^\circ$--twist defect in terms of the response
to normally incident plane waves.\footnote{It is worth  mentioning
that, although the  responses to normally incident plane waves are
the same for the two--section chiral STF satisfying \r{req} and
for a chiral STF with a central $90^\circ$--twist defect, the
$z$--directed components of the electric fields in the region
$0<z<D$ are different in general. Therefore, the two types of
devices, although functioning equivalently in terms of the
generation of spectral holes, are  not identical
electromagnetically.} Accordingly, both types of spectral holes
must emerge in the optical remittance spectrums  of a two--section
chiral STF obeying \r{req}   as the thickness $D$ changes; see the
spectrums of the co--polarized remittances in Figures \ref{Fig2} and \ref{Fig3} for
an illustrative example. The bandwidths of the spectral holes are
so small that dispersion of the constitutive scalars
$\eps_{a_j,b_j,c_j}$ can be ignored in most instances.

The theoretical underpinnings of the proposed method having been thus established,
let us turn our attention to the feasibility of fabricating the described device. Chiral STFs are fabricated
by directional physical vapor deposition, whereby the vapor of an inorganic material  is directed towards
a substrate at an angle ${ \chi^v}\in (0,\pi/2]$ to the substrate plane  \c{LM05,ML99}.
 Optical characterization experiments
on (nonchiral) columnar thin films \c{Hodg2} indicate that $\eps_{a,b,c}$ and $\chi$ are all monotonically increasing functions of
$\chi^v$. From the collected data, it has been shown that there exists a value $\chi^{pi}$ (called
the pseudoisotropic value) of $\chi$
such that $\eps_c=\tilde{\eps}_d$ \c{AL02}. Furthermore, $\eps_c \gtrless \tilde{\eps}_d$ for $\chi \gtrless \chi^{pi}$; thus, the local
birefringence changes sign as the pseudoisotropic value of $\chi$ is crossed.
The value of $\chi^{pi}$ is
dependent on the type of evaporant \c{Hodg2}   and most likely  on the deposition conditions as well \c{rm}.

It follows that the two sections of the proposed device must be deposited with vapors of different materials (or combinations
of materials). For example, suppose
$\tilde{\eps}_{d_1}>\eps_{c_1}$, and therefore
$\tilde{\eps}_{d_2}<{\eps}_{c_2}$. Then the section labeled $j=1$  must
be deposited at a low enough value of $\chi^v_1$ such that
$\chi_1< \chi_1^{pi}$, whereas the section labeled $j=2$ must
be deposited at a high enough value of $\chi^v_2$ such that
$\chi_2>\chi_2^{pi}$. Furthermore, the two materials
should be properly selected such that
$\eps_{c_1}+\tilde{\eps}_{d_1}=\eps_{c_2}+\tilde{\eps}_{d_2}$.

Could the proposed device be made by depositing vapor of a single material?
Based on the limited experimental data reported for
columnar thin films \c{Hodg2}, the answer is in the
negative. Since the sum $\eps_c+\tilde{\eps}_d$
increases monotonically with $\chi^v$, \r{req} cannot be fulfilled with just one material being deposited first with vapor directed
at angle $\chi^v_1$ and then at $\chi^v_2\ne \chi^v_1$.

Could a pair of CLCs be made to satisfy the conditions \r{req}?
The answer to this question is in the negative as well, because  (i)
$\chi=0$ for CLCs and (ii) the rodlike shapes of molecules \c{deG}
impose the restriction $\tilde{\eps}_d>\eps_c$. In other words,
neither a single--material, two--section chiral STF nor a combination of two CLCs can be utilized to
implement the proposed third method for the generation of the two
types of spectral holes. However, with the choice of a
single--material chiral STF for one section and a CLC for the
second section, it  could, in principle, be possible to  satisfy
\r{req}.

To conclude, we have theoretically established here the third
method (in addition to the ones calling for the insertion of layer
and twist defects) to generate circular--polarization--sensitive
spectral holes using STF technology. The proposed device is
optically similar to
 a chiral STF with a central
$90^\circ$--twist defect.  Finally, we have assessed the technological
feasibility of implementing the proposed method.\\

\bigskip

\noindent{\bf Acknowledgment}
This work was supported in part by a US National Science Foundation grant.
FW thanks Prof. J.A. Todd (Penn State) for continued support.

\end{document}